\documentclass[11pt]{article}
\usepackage{moriond,epsfig}

\def\Journal#1#2#3#4{{#1} {\bf #2}, #3 (#4)}

\def\PRL{\em Phys. Rev. Lett.}
\def\PRD{{\em Phys. Rev.} D}

\def\mco{\multicolumn}

\def\be{\begin{equation}}
\def\ee{\end{equation}}
\def\bea{\begin{eqnarray}}
\def\eea{\end{eqnarray}}

\begin{document}
\vspace*{4cm}
\title{THE CLEO-C PROJECT - A NEW FRONTIER OF OCQ PHYSICS}

\author{ HOLGER ST\"OCK }

\address{University of Florida, Department of Physics, PO Box 118440 \\
Gainesville, FL 32611-8440, USA}

\maketitle\abstracts
{
A proposal for a three-year program of charm and QCD physics with the CLEO 
detector operating in the range of $\sqrt{s}$ = 3 - 5 GeV is presented.
The CLEO-c program will include studies of semileptonic and hadronic charm
decays, as well as searches for gluonic matter in the area of nonpertubative
QCD. In addition, spectroscopy of the $\Upsilon$(1S), $\Upsilon$(2S) and
$\Upsilon$(3S) resonances is being carried out prior to the CLEO-c program.
}

\section{Introduction}
In this article the hadronic and QCD part of the proposed CLEO-c physics 
program will be outlined. The proposal includes a broad program of measurements
that will contribute to the understanding of important Standard Model processes
as well as provide the opportunity to probe the physics that lies beyond the
Standard Model. The dominant themes of this program are measurement of absolute
branching ratios for charm mesons with the precision of the order of 1 - 2\%
(depending upon the mode), determination of charm meson decay constants and of
the CKM matrix elements V$_{cs}$ and V$_{cd}$ at the 1 - 2\% level and 
investigation of processes in charm and $\tau$ decays, that are expected to be
highly suppressed within the Standard Model. Hence, a reconfigured CESR 
electron-positron collider operating at a center of mass energy range between
3 and 5 GeV together with the CLEO detector will give significant 
contributions to our understanding of fundamental Standard Model properties.

\section{Run Plan and Data Sets}
From the year 2003 to 2005 the CESR accelerator will be operated at
center-of-mass energies corresponding to $\sqrt{s} \sim$ 4140 MeV, $\sqrt{s}
\sim$ 3770 MeV ($\psi$'') and $\sqrt{s} \sim$ 3100 MeV (J/$\psi$). Taking
into account the anticipated luminosity which will range from 5$\times 10^{32}$
cm$^{-2}$s$^{-1}$ down to about 1$\times 10^{32}$ cm$^{-2}$s$^{-1}$ over this
energy range, the run plan will yield 3 fb$^{-1}$ each at the $\psi$'' and at
$\sqrt{s} \sim$ 4140 MeV above $D_s \bar{D_s}$ threshold and 1 fb$^{-1}$ at the
J/$\psi$. These integrated luminosities correspond to samples of 1.5 million
$D_s \bar{D_s}$ pairs, 30 million $D \bar{D}$ pairs and one billion J/$\psi$ 
decays. As a point of reference, these datasets will exceed those  of the
Mark III experiment by factors of 480, 310 and 170, respectively.
\par
In addition, prior to the conversion to low energy a total amount of 4 
fb$^{-1}$ spread over the $\Upsilon$(1S), $\Upsilon$(2S) and $\Upsilon$(3S) 
resonances is taken to launch the QCD part of the program. These data sets will
increase the available $b\bar{b}$ bound state data by more than an order of
magnitude.

\section{Hardware Requirements}
The conversion of the CESR accelerator for low energy operation will require
the addition of 18 meters of wiggler magnets to enhance transverse cooling of
the beam at low energies. In the CLEO III detector the solenoidal field will be
reduced to 1.0 T and the silicon vertex detector will be replaced with a small,
low mass inner drift chamber. No other requirements are necessary.

\section{Physics Program}
The following sections will outline the proposed CLEO-c physics program.
Section \ref{sec:ypsilon} will focus on the Ypsilon spectroscopy, section
\ref{sec:charm} will describe the charm decay program and finally section
\ref{sec:glue} will give an overview about the gluonic matter studies.

\subsection{Ypsilon Spectroscopy \label{sec:ypsilon}}
From fall 2001 to summer 2002 CLEO will collect data on the $\Upsilon$ 
resonances below the $\Upsilon$(4S). During the resonance running CLEO 
anticipates in excess of 4 fb$^{-1}$ of accumulated data.
\par
So far, the only established states below $B\bar{B}$ threshold are the three
vector singlet $\Upsilon$ resonances ($^3S_1$) and the six $\chi_b$ and
$\chi'_b$ (two triplets of $^3P_J$) that are accessible from these parent
vectors via E1 radiative transitions. By collecting substantial data samples
at the $\Upsilon$(1S), $\Upsilon$(2S) and $\Upsilon$(3S), CLEO will address a
variety of outstanding physics issues.

\begin{itemize}

\item Discovery of $\eta_b$ and Observation of h$_b$ \newline
The $\eta_b$ is the ground state of $b\bar{b}$. Most present theories indicate
the best approach would be the hindered M1 transition from the $\Upsilon$(3S),
with which CLEO might have a signal of 5 $\sigma$ significance in 1 fb$^{-1}$
of data. In the case of the h$_b$, CLEO established an upper limit of
$\cal{B}$($\Upsilon$(3S)$\rightarrow \pi^+ \pi^- $h$_b$) $<$ 0.18\% at 90\%
confidence level \cite{hb}. This result, based on $\sim$ 110 pb$^{-1}$, already
tests the theoretical predictions \cite{hb2} for this transition which range
from 0.1 - 1.0\%. The resonance run program will measure the mass of the h$_b$,
assuming the predictions are valid, to $\sim$ 5 MeV.

\item Observation of 1$^3$D$_J$ states \newline
The $b\bar{b}$ system is unique as it has states with L = 2 that lie below
the open-flavor threshold. These states have been of considerable theoretical
interest, as indicated by many predictions of the center-of-gravity of the
triplet. Based on both theoretical estimates and previous CLEO data samples,
CLEO should see 20 - 40 events in the four-photon cascade
$\Upsilon$(3S)$\rightarrow \gamma_1\chi'_b
\rightarrow \gamma_1\gamma_2(^3$D$_J) \rightarrow \gamma_1\gamma_2\gamma_3
\chi_b \rightarrow \gamma_1\gamma_2\gamma_3\gamma_4\ell^+\ell^-$

\item Search for glueball candidates in radiative $\Upsilon$(1S) decays \newline
The BES collaboration has reported signals for a glueball candidate 
\cite{radiative} in radiative J/$\psi$ decay - a glue-rich environment. Naively
one would expect the exclusive radiative decay to be suppressed in $\Upsilon$
decay by a factor of roughly 40, which implies product branching fractions for
$\Upsilon$ radiative decay of $\sim$ 10$^{-6}$. With 1 fb$^{-1}$ of data and
efficiencies of around 30\% one can expect $\sim$ 10 events in each of the
exclusive channels, which would be an important confirmation of the J/$\psi$
studies.

\end{itemize}

\subsection{Charm Decays \label{sec:charm}}
The observable properties of the charm mesons are determined by the strong
and weak interactions. As a result, charm mesons can be used as a laboratory 
for the studies of these two fundamental forces. Threshold charm experiments
permit a series of measurements that enable direct study of the weak
interactions of the charm quark as well as tests of our theoretical technology
for handling the strong interactions.

\subsubsection{Semileptonic Charm Decays}
The CLEO-c program will provide a large set of precision measurements in the 
charm sector against which the theoretical tools needed to extract CKM matrix
information precisely from heavy quark decay measurements will be tested and
honed.
\par
CLEO-c will measure the branching ratios of many exclusive semileptonic modes,
including
$D^0 \rightarrow K^- e^+ \nu$,
$D^0 \rightarrow \pi^- e^+ \nu$,
$D^0 \rightarrow K^{-} e^+ \nu$,
$D^+ \rightarrow \bar{K}^{0} e^+ \nu$,
$D^+ \rightarrow \pi^0 e^+ \nu$,
$D^+ \rightarrow \bar{K}^{0*} e^+ \nu$,
$D^+_s \rightarrow \phi e^+ \nu$ and
$D^+_s \rightarrow \bar{K}^{0*} e^+ \nu$.
The measurement in each case is based on the use of tagged events where the
cleanliness of the environment provides nearly background-free signal samples,
and will lead to the determination of the CKM matrix elements
$\mid$V$_{cs}$$\mid$ and $\mid$V$_{cd}$$\mid$ with a precision level of 
1.6\% and 1.7\%, respectively. Measurements of the vector and axial vector form
factors V($q^2$), A$_1$($q^2$) and A$_2$($q^2$) will also be possible at the
$\sim$ 5\% level. Table \ref{tab:semileptonic} summarizes the proposed branching
fractional errors.
\par
HQET provides a successful description of the lifetimes of charm hadrons and
of the absolute semileptonic branching ratios of the D$^0$ and D$_s$ 
\cite{hqe}. Isospin invariances of the strong forces lead to corrections of
$\Gamma_{SL}$(D$^0$) $\simeq \Gamma_{SL}$(D$^+$) in the order of 
$\cal{O}$(tan$^2 \Theta_C$) $\simeq$ 0.05. Likewise SU(3)$_{Fl}$ symmetry
relates $\Gamma_{SL}$(D$^0$) and $\Gamma_{SL}$(D$^+$), but a priori would
allow them to differ by as much as 30\%. However, HQET suggests, that they 
should agree to within a few percent. A charm factory is the best place to
measure absolute inclusive semileptonic charm branching ratios, in particular
$\cal{B}$(D$_s \rightarrow$ X$\ell\nu$) and thus $\Gamma_{SL}$(D$_s$).

\subsubsection{Hadronic Charm Decays}
Of the charm hadrons, the D$^0$ is the best known. The CLEO and ALEPH 
experiments provide by far the most precise measurements for the decay
D$^0 \rightarrow$ K$^- \pi^+$. They use the same technique, where they look at
$D^{*+} \rightarrow \pi^+ D^0$ decays and take the ratio of the D$^0$ decays
into K$^- \pi^+$ to the number of decays with only the $\pi^+$ from the D$^{*+}$
decay detected. The dominant systematic uncertainty is the background level in
the latter sample. In both experiments, the systematic errors exceed the
statistical errors. By using $D^0\bar{D}^0$ decays and tagging both D mesons
the background can be reduced to almost zero and the branching ratio fractional
error can be improved significantly (see Table \ref{tab:hadronic}).
\par
The D$^+$ absolute branching ratios are determined by using fully reconstructed 
D$^{*+}$ decays, comparing $\pi^0$ D$^+$ with $\pi^+$ D$^0$ and using isotropic
spin symmetry. Hence, this rate cannot be determined any better than the
absolute D$^0$ decay rate using this technique. By using D$^+$D$^-$ decays and
a double tag technique the background can be reduced again to almost zero
which leads to a significant improvement of the branching ratio fractional 
error (see Table \ref{tab:hadronic}).

\begin{table}[t]
\caption{Proposed branching fractional errors for semileptonic decay modes
\label{tab:semileptonic}}
\vspace{0.4cm}
\begin{center}
\begin{tabular}{|l|c|c|}
\hline
Decay Mode                      & \mco{2}{|c|}{BR fractional error \%} \\
\cline{2-3}
                                & PDG 2000 & CLEO-c                    \\
\hline
$D^0 \rightarrow K \ell \nu$    & 5        & 1.6                       \\
$D^0 \rightarrow \pi \ell \nu$  & 16       & 1.7                       \\
$D^+ \rightarrow \pi \ell \nu$  & 48       & 1.8                       \\
$D_s \rightarrow \phi \ell \nu$ & 25       & 2.8                       \\
\hline
\end{tabular}
\end{center}
\end{table}

\begin{table}[t]
\caption{Proposed branching fractional errors for hadronic decay modes
\label{tab:hadronic}}
\vspace{0.4cm}
\begin{center}
\begin{tabular}{|l|c|c|}
\hline
Decay Mode                 & \mco{2}{|c|}{BR fractional error \%} \\
\cline{2-3}
                           & PDG 2000 & CLEO-c                    \\
\hline
$D^0 \rightarrow K \pi$    & 2.4      & 0.5                       \\
$D^+ \rightarrow K K \pi$  & 7.2      & 1.5                       \\
$D_s \rightarrow \phi \pi$ & 25       & 1.9                       \\
\hline
\end{tabular}
\end{center}
\end{table}

\subsection{Gluonic Matter \label{sec:glue}}
With approximately one billion J/$\psi$ produced, CLEO-c will be the natural
glue factory to search for glueballs and other glue-rich states using J/$\psi$
$\rightarrow$ gg $\rightarrow \gamma$X decays. The region of 
\linebreak
1 $<$ M$_X$ $<$ 3 GeV/c$^2$ will be explored with partial wave analyses for
evidence of scalar or tensor glueballs, glueball-$q\bar{q}$ mixtures, exotic
quantum numbers, quark-glue hybrids and other new forms of matter predicted by
QCD. This includes the establishment of masses, widths, spin-parity quantum
numbers, decay modes and production mechanisms for any identified states, an in
detail exploration of reported glueball candidates such as the tensor candidate
f$_J$(2220) and the scalar states f$_0$(1370), f$_0$(1500) and f$_J$(1710), and
the examination of the inclusive photon spectrum J/$\psi \rightarrow \gamma$X 
with $<$ 20 MeV photon resolution and identification of states with up to 100
MeV width and inclusive branching ratios above 1$\times 10^{-4}$. In addition
spectroscopic searches for new states of the $b\bar{b}$ system and for exotic
hybrid states such as $cg\bar{c}$ will be made using the 4 fb$^{-1}$
$\Upsilon$(1S), $\Upsilon$(2S) and $\Upsilon$(3S) data sets. Analysis of
$\Upsilon$(1S) $\rightarrow \gamma$X will play an important role in verifying
any glueball candidates found in the J/$\psi$ data.

\section{Summary}
The high-precision charm and quarkonium data will permit a broad suite of
studies of weak and strong interaction physics. In the threshold charm sector
measurements are uniquely clean and make possible the unambigous determinations
of physical quantities discussed above. CLEO-c will utilize a variety of tools,
namely J/$\psi$ radiative decays, two-photon collisions (using almost real, as
well as highly virtual space-like photons), deep inelastic Coulomb scattering 
and continuum production via $e^+ e^-$ annihilation to obtain significant new
information on the spectrum of hadrons, both normal and exotic, and their decay
channels. A quantitative improvement can be expected not only from the large
accumulated statistics, but also from combining the results obtained using all
these tools together with the results from the $\Upsilon$ resonance runs. The
significance of this is better sensitivity, reduced systematics and a better
chance to obtain a coherent picture of the hadron sector.

\section*{Acknowledgments}
I am delighted to acknowledge the invaluable contributions of many individuals
to the development of the CLEO-c and CESR-c program and the outstanding
contributions of my CLEO colleagues over the life of the experiment. The
experimental aspects of this program are based on their effort and experience.


\end{document}